\documentstyle[11pt,paspconf]{article}

\include{psfig} \psnoisy 

\begin{document}

\title{What are the Galaxies Contributing to the Cosmic Infrared Background~?}
\author{B. Guiderdoni}
\affil{Institut d'Astrophysique de Paris, 98bis Boulevard Arago, 
F--75014 Paris, France}

\begin{abstract}
Recent optical observations have led to a
significant progress in our understanding of galaxy formation and evolution.
However, our view on the deep universe is currently limited to the
starlight which directly escapes from high--redshift 
galaxies, since we so far ignore the fraction of luminosity
absorbed by dust and released in the IR/submm wavelength range.
A new constraint is set by the possible detection of the
Cosmic Infrared Background. We briefly review the observations and
use a semi--analytic model of galaxy formation and evolution
to predict number counts consistent with the level of the background.
It turns out that the predictions fairly accomodate preliminary data
at 175 and 850 $\mu$m. This suggests that a significant fraction of
star/galaxy formation at high $z$ is hidden by dust.
\end{abstract}


\section{Introduction}

The picture of galaxy formation and evolution which is progressively 
emerging from deep optical surveys is fascinating. The
measurement of the cosmic star formation rate 
(hereafter SFR) density of the universe up to $z \sim 4$
from rest--frame UV fluxes seems to show a strong peak at $z \sim 1.5$
and suggests that we could have seen the bulk
of star formation in the universe
(Lilly {\it et al.} 1996; Madau {\it et al.} 1996). 
This high SFR seems to be correlated with
the decrease of the cold--gas comoving
density in damped Lyman--$\alpha$ systems
between $z=2$ and $z=0$ (Storrie--Lombardi {\it et al.} 1996). 
These results nicely fit in 
a view where star formation in bursts triggered by interaction/merging
consumes and enriches the gas content of galaxies as time goes on.
Indeed, such a scenario is qualitatively predicted within the paradigm of 
hierarchical growth of structures in which galaxy formation is a continuous
process.

However, we have only a partial view on galaxy 
evolution since most of the observational data come from optical
surveys which probe the rest--frame UV and visible emission of 
high--$z$ galaxies.
A still unknown fraction of star/galaxy formation is hidden by dust which
absorbs UV/visible starlight and re--radiates at larger wavelengths.
At low $z$, the IRAS satellite has discovered a sequence of IR 
properties, from ``normal spirals'' to the ``luminous IR galaxies''
(hereafter LIRGs), mostly interacting systems, and the
spectacular ``ultraluminous IR galaxies'' (hereafter ULIRGs), 
which are mergers 
(Sanders and Mirabel 1996) and emit more than 95 \% of their
energy in the IR. 

\begin{figure}[htbp]
\vbox 
{\vskip -0.8truecm
\psfig{figure=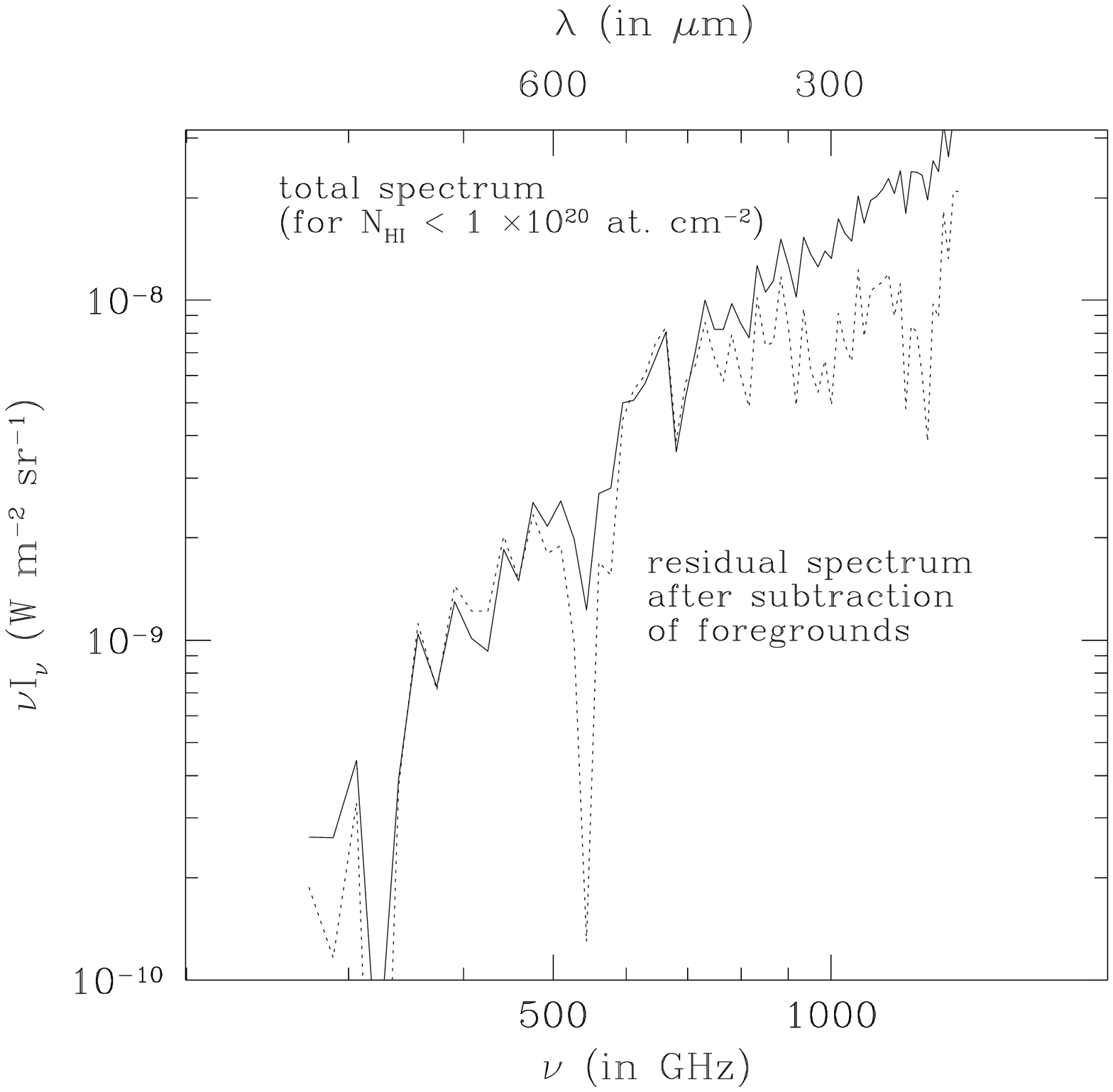,width=0.7\textwidth}
\vskip -0.8truecm
\caption{FIRAS residuals in the cleanest regions of the sky
(solid line), and the CIRB (dotted line).}
\psfig{figure=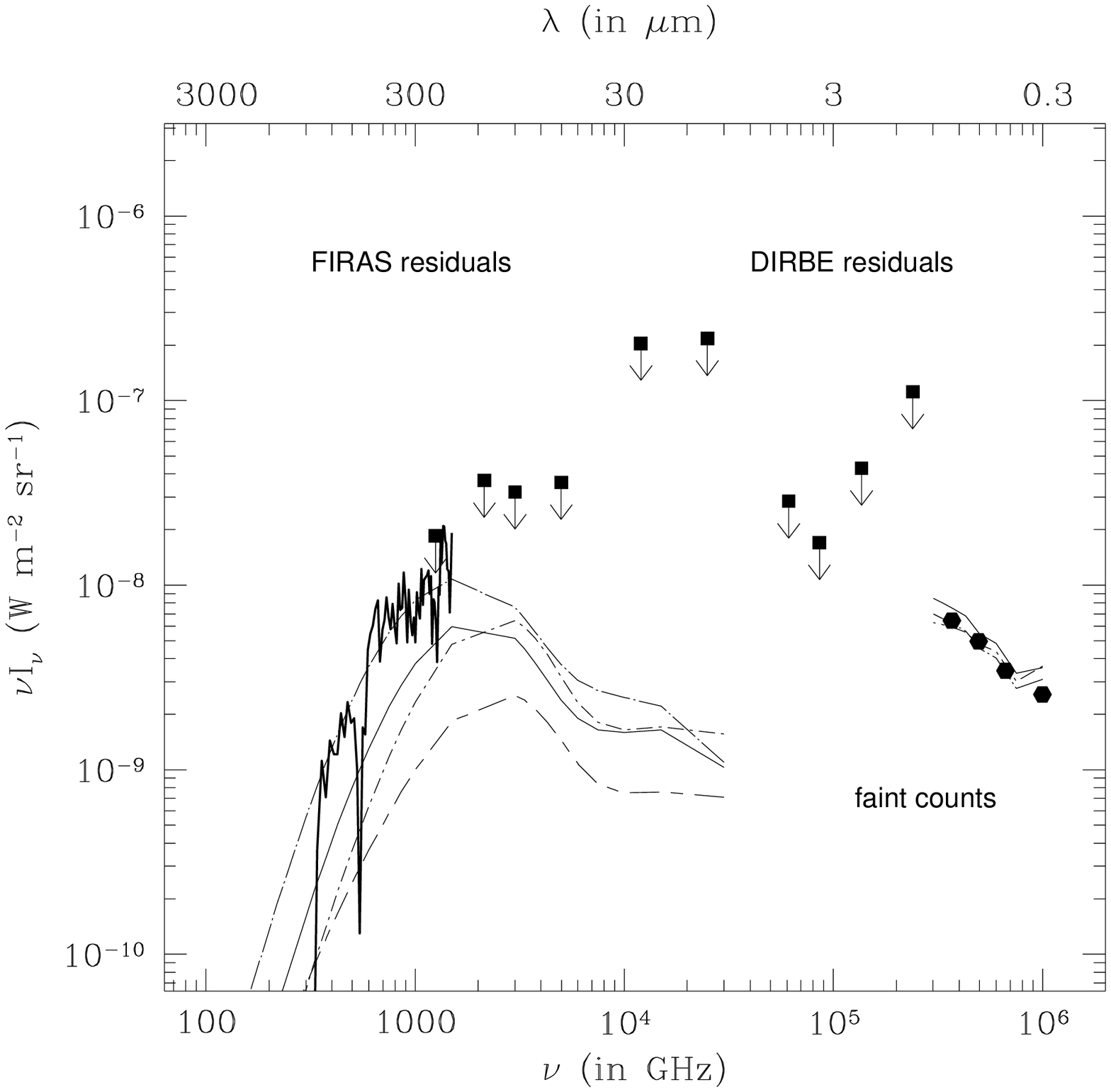,width=0.7\textwidth}
\vskip -1truecm
\caption{Diffuse backgrounds in the FIR/submm and in the optical. Solid
triangles: COBE/DIRBE upper limits (Hauser 1995).
Thick solid line: CIRB (Puget {\it et al.}
1996; Guiderdoni {\it et al.} 1997a). Solid hexagons:
Cosmic Optical Background from faint galaxy counts (Williams {\it et al.}
1996). Long--dashes--and--short--dashes: predictions
without evolution. Dots--and--short--dashes: quiescent star formaiton
history. Solid line : scenario
A without ULIRGs. Dots--and--long--dashes: scenario E with ULIRGs, fitting
the CIRB. See Guiderdoni {\it et al.} (1997b) for details.}
}
\end{figure}

In contrast, we know very little
about galaxy evolution at high $z$ in this wavelength range. 
Faint galaxy counts with IRAS seem to show a strong luminosity and/or
density evolution at $z \la 0.2$, but it is difficult to extrapolate
this trend to higher redshifts on a firm ground. 
The Cosmic Infrared Background (hereafter CIRB) detected by Puget {\it et al.}
(1996) in FIRAS residuals sets strong constraints on the presence of a 
population of IR/submm sources. We hereafter briefly review the observational 
case (in Sec. 2) 
and use a semi-analytic modelling of galaxy formation and evolution
to predict number counts (in Sec. 3).

\section{The diffuse background due to galaxies}

The epoch of galaxy formation can be observed in
the background radiation which is produced by the accumulation
of the light of extragalactic sources along the line of sight. 
The search for the ``Cosmic Optical Background''
currently gives only upper limits. The shallowing
of the HDF faint counts suggests that we are now close to
convergence and that an estimate of the COB can 
be obtained by summing up the contributions of faint galaxies
(Williams {\it et al.} 1996). 

\begin{center}
\begin{figure}[htbp] 
\psfig{figure=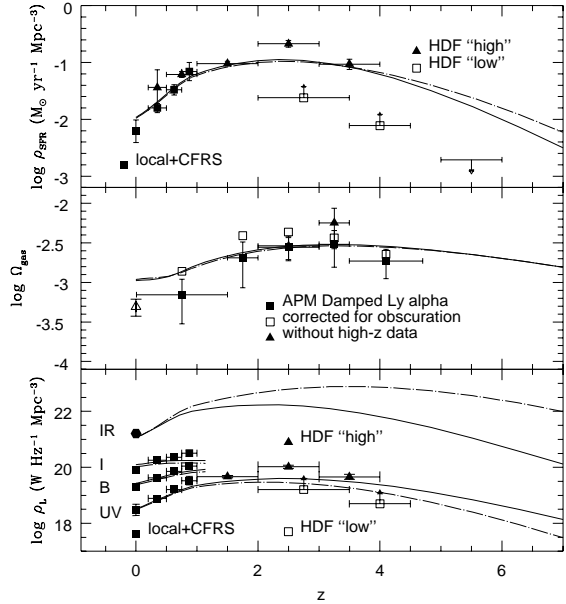,width=0.7\textwidth}
\vskip -1truecm
\caption{Evolution of the ``cosmic constraints'' for scenario A (solid line)
without ULIRGs and scenario E (dots and long dashes) with a fraction of ULIRGs 
increasing with $z$. {\it Upper panel:}
comoving star formation rate density as
computed from rest--frame UV luminosity densities, by using Salpeter IMF with
slope $1.35$ {\it without} extinction. 
{\it Middle panel:} cold gas density parameter in damped 
Lyman--$\alpha$ absorbers. {\it Lower panel:} rest--frame luminosity
densities.}
\end{figure}
\end{center}

The re--analysis of COBE/FIRAS residuals
between 200 $\mu$m and 2 mm has led Puget {\it et al.} (1996) to
discover the presence of an isotropic component which is likely to
be the long--sought CIRB.
Several steps are necessary in order to
remove the foregound Galactic components and extract the isotropic residual
identified as the CIRB. Whereas, at the wavelengths probed by FIRAS, the
interplanetary emission is small and easily removed, the
emission from interstellar dust mixed with the different gas phases of the
interstellar medium is the dominant component. In order to
address this problem, the original method of Puget {\it et al.} 
was applied again by Guiderdoni {\it et al.} (1997a), 
but only in the cleanest regions with very low HI column densities
($N_{HI} \leq 1 \times 10^{20}$ atoms cm$^{-2}$ instead of
$N_{HI} \leq 4.5 \times 10^{20}$ atoms cm$^{-2}$). In that case, the residual
component totally dominates the emission, as shown in Fig. 1.  
Fig. 2 shows that the CIRB intensity per frequency decade $\nu
I_{\nu} \simeq (7 \pm 0.3)~10^{-9}$ W m$^{-2}$ sr$^{-1}$ near 
300 $\mu$m is a factor of 5
higher than the no--evolution prediction obtained by a simple extrapolation of
the IR luminosities of local galaxies.  

Such a detection yields the first 
``post--IRAS'' constraint on the high--$z$ 
evolution of galaxies in the IR/submm range,
before the era of ISO results. Its level is comparable to the
optical background estimated by summing up faint galaxy counts down
to the deepest limit so far available which is given by the Hubble Deep Field
(Williams {\it et al.} 1996), and suggests
that a significant fraction of the energy of young stars is absorbed by
dust and released in the IR/submm.

The DIRBE instrument on COBE has given so far
upper limits on the IR background at wavelengths between 2 and 300 $\mu$m
(Hauser 1995). However Schlegel {\it et al.} (1997) extract a uniform 
background radiation from DIRBE observations at the levels $\nu
I_{\nu} \simeq (17 \pm 4)~10^{-9}$ W m$^{-2}$ sr$^{-1}$ at 240 $\mu$m and $\nu
I_{\nu} \simeq (32 \pm 13)~10^{-9}$ W m$^{-2}$ sr$^{-1}$ at 
140 $\mu$m. At least
the value at 240 $\mu$m, where the foregrounds are less important, seems
to be consistent with the estimate from FIRAS. A more thorough analysis of the
DIRBE data is currently undertaken by the DIRBE team.

\section{Semi--analytic modelling}

In order to 
analyse the current data and make predictions for forthcoming observations,
we have chosen to model galaxy evolution by
using the so--called ``semi--analytic'' approach which
has been rather successful in reproducing the overall
properties of galaxies in the optical range. We have elaborated an
extension of this type of method to the IR/submm range. Details of the
modelling, and extensive predictions are 
given elsewhere (Guiderdoni {\it et al.} 1997a,b). 
In the Standard Cold Dark Matter cosmological
scenario ($H_0=50$ km s$^{-1}$ Mpc$^{-1}$, $\Omega_0=1$, $\Lambda=0$,
$\Omega_b=0.05$, $\sigma_8=0.67$), we have designed a family of plausible
evolutionary scenarios with different fractions of ULIRGs. In the following,
scenario A has no ULIRGs and scenario E has a fraction of
ULIRGs increasing with $z$.

Fig. 3 shows the predicted comoving SFR, gas, and luminosity
densities in the universe, for these
scenarios. The strong episode of star formation at $z \sim 1$
corresponds to the decrease of the gas density in the universe.
The local luminosity densities in the UV, visible and IR are
accommodated by all our evolutionary scenarios, though they predict 
high--$z$ IR luminosity densities which are strongly different.
 The predicted backgrounds generated with our scenarios are displayed
in Fig. 2. Scenario E nicely fits the level and shape of the CIRB. 

\begin{center}
\begin{figure}[htbp] 
\psfig{figure=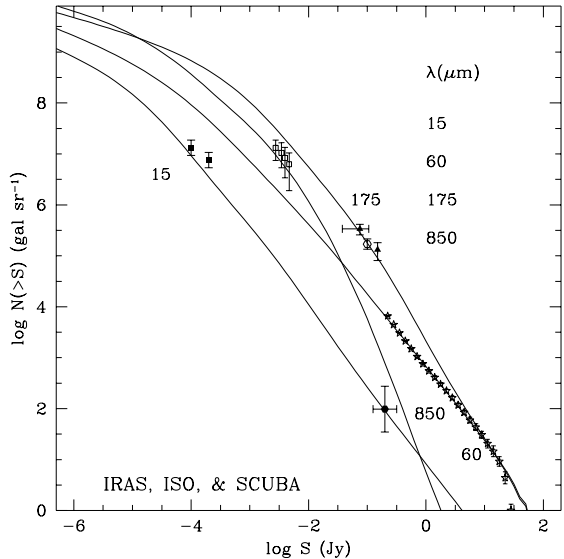,width=0.6\textwidth}
\caption{Predictions for faint galaxy counts at 15 $\mu$m, 
60 $\mu$m, 175 $\mu$m, and 850 $\mu$m for scenario E fitting the CIRB.
See text for the keys of the symbols.}
\end{figure}
\end{center}

The correponding 
predictions for faint galaxy counts are given in Fig. 4, with data from
IRAS counts at 60 $\mu$m (Lonsdale {\it et al.} 1990; open stars), 
the ISO--HDF follow--up with ISOCAM 
at 15 $\mu$m (Oliver {\it et al.} 1997; solid squares) to be compared
with data from the IRAS survey (Rush {\it et al.} 1993; solid dot), 
preliminary data for the
Lockman Hole (Kawara {\it et al.} 1998, in preparation; open dot) 
and the Marano field (Puget {\it et al.} 1998, in preparation; solid
triangles)
with ISOPHOT at 175 $\mu$m, and the first deep field observed with 
the JCMT/SCUBA at 850 $\mu$m (Smail {\it et al.} 1997; open squares). 
The agreement
of the predictions with the data seems good enough to suggest that 
these counts do probe the evolving 
population contributing to the CIRB. The model
shows that 15 \%  and 60 \% of the CIRB respectively at 175 $\mu$m 
and 850 $\mu$m are built up by objects brighter than
the current limits of ISOPHOT and
SCUBA deep fields. It is also possible to get predicted redshift 
distributions for these counts. The predicted median redshift of the ISO--HDF
is $z \sim 0.8$. It increases to $z \sim 1.5$ for the deep ISOPHOT
surveys, and to $z \ga  2$ for SCUBA, even if the latter value seems to be
very sensitive to the details of evolution.

\section{Conclusions}

With ISO and SCUBA, a new window is now open to deep extragalactic 
surveys. A large number of 
objects are expected from current models of galaxy evolution in the 
IR/submm. These models accommodate local IR data and follow
the high--redshift evolution of the ``cosmic constraints'' (that is, the
comoving SFR, gas and UV/optical/NIR luminosity densities), as well as
the high level of the isotropic component in COBE residuals 
which is probably the CIRB. The data seem to be consistent with a strong 
evolution. Significant progress in resolution 
and/or sensitivity expected from forthcoming satellites such as WIRE, 
SIRTF, and FIRST will show the high--redshift counterparts of 
local ULIRGs.
The ``discovery potential'' of
the forthcoming IR/submm observations is very exciting.

\acknowledgments

I thank my collaborators F.R. Bouchet, E. Hivon, G. Lagache, 
B. Maffei, and J.L. Puget.

\end{document}